\documentclass[aps,showpacs,twocolumn,floats,superscriptaddress]{revtex4}
\usepackage[latin9]{inputenc}
\usepackage{amsmath}
\usepackage{amssymb}
\usepackage{graphicx}
\makeatletter

\providecommand{\tabularnewline}{\\}

\@ifundefined{textcolor}{}
{%
 \definecolor{BLACK}{gray}{0}
 \definecolor{WHITE}{gray}{1}
 \definecolor{RED}{rgb}{1,0,0}
 \definecolor{GREEN}{rgb}{0,1,0}
 \definecolor{BLUE}{rgb}{0,0,1}
 \definecolor{CYAN}{cmyk}{1,0,0,0}
 \definecolor{MAGENTA}{cmyk}{0,1,0,0}
 \definecolor{YELLOW}{cmyk}{0,0,1,0}
}

\usepackage{color}%\bibliographystyle{plain}
\def\NOT(#1,#2){\OneQubitGate(#1,#2){$X$}}

\makeatother

\begin{document}

\title{Experimental protection of 2-qubit quantum gates against environmental
noise by dynamical decoupling }

\author{Jingfu Zhang and Dieter Suter}

\affiliation{Fakultät Physik, Technische Universität Dortmund, D-44221 Dortmund,
Germany }

\date{\today}
\begin{abstract}
Hybrid systems consisting of different types of qubits are promising
for building quantum computers if they combine useful properties of
their constituent qubits. However, they also pose additional challenges
if one type of qubits is more susceptible to environmental noise than
the others. Dynamical decoupling can help to protect such systems
by reducing the decoherence due to the environmental noise, but the
protection must be designed such that it does not interfere with the
control fields driving the logical operations. Here, we test such
a protection scheme on a quantum register consisting of the electronic
and nuclear spins of a nitrogen-vacancy center in diamond. The results
show that processing is compatible with protection: The dephasing
time was extended almost to the limit given by the longitudinal relaxation
time of the electron spin.
\end{abstract}


\maketitle
\textit{--Introduction.} Choosing the most suitable qubit systems
and implementing high-precision quantum gate operations on them may
be considered as the main challenges for building quantum computers
that will be capable of solving certain problems qualitatively faster
than classical computers \cite{nielsen,Stolze:2008xy}. In every physical
realization, environmental noise is present, which degrades the information
stored in the quantum systems. Several techniques have been developed
for protecting quantum systems against this effect, including passive
schemes like decoherence-free subspaces \cite{3045} or active techniques
like dynamical decoupling (DD) \cite{5003,suter2012,Paolo199977}.
So far, these techniques have been applied mostly to the protection
of quantum states in quantum memories \cite{liu2009,hanson2010,cory2011,PhysRevLett.106.040501,nature_514_72,shim12}.
However, similar protection is also required for quantum information
processing. Here, the protection must be designed in such a way that
it decouples the protected system from environmental noise, but not
from the control fields that are applied for driving the gate operations.
Experimental implementations of protected single-qubit gates and some
special 2-qubit gates, such as a controlled-NOT (CNOT) gate, were
recently reported \cite{prar2012,PhysRevLett.112.050502,dobrovitski2012,Natcomm3254},
but so far no scheme was demonstrated that can be applied to arbitrary
two-qubit gates in different physical qubit systems.

Implementing quantum computing in hybrid systems is a promising strategy
for building quantum computers, as it increases the range of possible
systems and allows one to combine useful properties of different types
of qubits ~\cite{PhysRevLett.105.140503,PhysRevLett.105.140502,PhysRevLett.105.140501,PhysRevB.82.024413}.
As a typical hybrid system, the nitrogen vacancy (NV) center in diamond
consists of an electronic spin $S=1$ and a nitrogen nuclear spin
$I=1$ and combines several useful properties for a quantum register.
It also poses some challenges, mostly because the characteristic properties
of the two types of spins differ by many orders of magnitude. For
example, the natural coherence time at room temperature of the electron
spin is of the order of 1 $\mu$s, while that of the nuclear spin
is $5$ ms \cite{dobrovitski2012}. The electron spin dephasing time
is usually shorter than typical durations of gates applied to the
nitrogen spin. The critical gate operations in such a system are therefore
the two-qubit gates in which the electron spin is the control quit,
while the nuclear spin is the target qubit: In this situation, the
relatively slow Rabi frequency of the nuclear spin results in a long
duration of the gate, during which the whole quantum information must
be conserved, including superpositions of different basis states of
the electron spin.

It is possible to apply DD pulses that act parallel to the control
field driving the targeted nuclear spin \cite{dobrovitski2012}. However,
since the DD pulses interchange the computational basis states of
the control qubit (the electron spin), the segments of the control
field separated by the DD pulses implement two gates with different
control conditions. Obtaining the targeted evolution requires therefore
careful synchronization of the DD pulses with the internal system
evolution. In some systems, specific properties of the Hamiltonian
allow one to design DD sequences such that they protect the system
from environmental noise and simultaneously implement a useful gate
operation ~\cite{Natcomm3254}. However, this approach is very specific
to the system and requires that the quantum system can be simulated
classically. It is therefore difficult to extend this approach to
large quantum registers.

Here, we present a strategy that avoids these limitations and can
be applied to arbitrary gate operations with minimal overhead. We
demonstrate the scheme for the case of a 2-qubit system, where the
duration of one gate is long enough that decoherence destroys a significant
part of the quantum information, unless DD is applied to protect it.
Compared to existing schemes, our approach can implement arbitrary
2-qubit gates and does not require fine-tuning to system parameters.
It requires that the control field driving the gate operation is split
into serval segments that can be inserted into the delays of the DD
sequence. The segments have to be adjusted to take the effect of the
DD pulses into account, which interchange the computational basis
states $|0\rangle\leftrightarrow|1\rangle$ and therefore the condition
for the controlled 2-qubit gate.

\textit{--Experimental protocol and results.} We illustrate this basic
idea for the case of a controlled rotation, CR. The unitary operation
for this gate can be represented as
\begin{equation}
U_{CR}(\theta)=\left(\begin{array}{cc}
\mathbf{1} & \mathbf{0}\\
\mathbf{0} & R_{x}(\theta)
\end{array}\right),\label{UCRmat}
\end{equation}
where $\mathbf{1}$, $\mathbf{0}$ and and $R_{x}(\theta)$ represent
$2\times2$ matrices corresponding to the unit operator, the zero
matrix and a rotation matrix around the $x$-axis, $R_{x}(\theta)=e^{-i\theta\sigma_{x}/2}$
where $\sigma_{x}$ denotes the $x$-component of the Pauli matrix.

\begin{figure}
\centering{}\includegraphics[width=1\columnwidth]{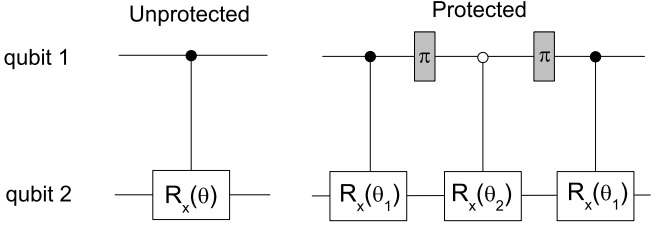}
\protect\protect\protect\caption{Circuits for the
controlled-rotation gate and its protected version.
$R_{x}(\theta)$ represents a rotation around the $x$-axis by an
angle $\theta$. The rotation angles are related by
$2\theta_{1}+\theta_{2}=\theta$. The filled rectangles are the DD
pulses with flip angle $\pi$. \label{figcirCNOT}}
\end{figure}

Fig.\ref{figcirCNOT} shows the circuit models of the CR gates. The
left-hand part represents the unprotected version, while the right-hand
side represents the protected gate for the simplest DD sequence consisting
of two refocusing pulses. Filled circles indicate that the operation
is performed on the subspace where the control qubit is in the logical
state $|1\rangle$, empty circles stand for the control condition
$|0\rangle$. When $\theta=\pi$, $U_{CR}$ becomes the CNOT gate,
with an additional phase gate $e^{-i\pi\sigma_{z}/4}$ on qubit 1
and an overall phase.

\begin{figure}
\centering{}\includegraphics[width=1\columnwidth]{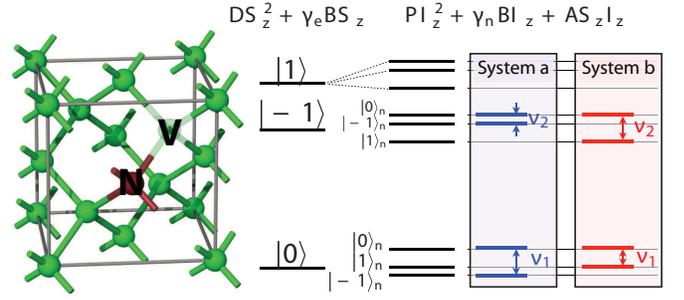}
\protect\protect\protect\caption{(color online). Structure and energy level scheme of the NV center.
Two subspaces of the total system, which are used in the experiment,
are shown and marked as subsystems $a$ and $b$. The quantum numbers
refer to the electronic spin $S$ and the nuclear spin $I$, respectively.
\label{figlevels}}
\end{figure}

For the experimental implementation, we chose a 4-level system out
of the 9-level system composed by the electronic and nuclear spins
of the NV center in diamond shown in Fig. \ref{figlevels}. If the
magnetic field is oriented along the NV symmetry axis, the Hamiltonian
of the NV center can be written as ~\cite{arXiv:1405.2696}
\begin{equation}
H_{s}=DS_{z}^{2}+\gamma_{e}BS_{z}+PI_{z}^{2}+\gamma_{n}BI_{z}+AS_{z}I_{z}.\label{Hamsim}
\end{equation}
Here $S_{z}$ and $I_{z}$ are the $z$-components of the spin-1 operators
for the electron and nitrogen ($^{14}$N) nuclear spins, respectively.
The zero-field splitting is $D=2.87$ GHz, the nuclear quadrupolar
splitting is $P=-4.95$ MHz, and the hyperfine coupling is $A=2.16$
MHz ~\cite{PhysRevB.89.205202,PhysRevB.47.8816}. The electronic
gyromagnetic ratio is $\gamma_{e}=2.8$ MHz/G, and the nuclear gyromagnetic
ratio $\gamma_{n}=0.30$ kHz/G. The strength of the field in our experiment
was about $87$ G.

In the 9-dimensional Hilbert space of the electronic and nuclear spins,
we chose two different 2-qubit systems to implement the gates. In
Fig. \ref{figlevels}, they are denoted as $a$ and $b$, respectively.
In each system, we use the electron spin as the control qubit (qubit
1), while the nuclear spin of the nitrogen represents the target qubit
(qubit 2). The logical states for the two qubits correspond to
\begin{eqnarray}
\{|0\rangle_{1},|1\rangle_{1}\} & = & \{|0\rangle_{e},|-1\rangle_{e}\}\nonumber \\
\{|0\rangle_{2},|1\rangle_{2}\} & = & \{|0\rangle_{n},|\mp1\rangle_{n}\},\label{eq:CompBase}
\end{eqnarray}
where the minus sign refers to system $a$ and the plus sign to
system $b$. Table \ref{tableNparams} lists the measured transition
frequencies of the nuclear spin and the corresponding Rabi
frequencies.

%%%%------------- table -------------------------------------------------------%%
\begin{table}
\begin{tabular}{|c|c|c|}
\hline
 & Transition  & Rabi \tabularnewline
 & frequency (MHz)  & frequency (kHz) \tabularnewline
\hline
System a  & $\nu_{1}=4.970$  & $\Omega_{1}/(2\pi)=9.05$ \tabularnewline
 & $\nu_{2}=2.808$  & $\Omega_{2}/(2\pi)=5.84$ \tabularnewline
\hline
System b  & $\nu_{1}=4.918$  & $\Omega_{1}/(2\pi)=9.00$ \tabularnewline
 & $\nu_{2}=7.088$  & $\Omega_{2}/(2\pi)=5.60$ \tabularnewline
\hline
\end{tabular}\protect\protect\protect\protect\protect\protect\protect\protect\protect\caption{Measured parameters of the nuclear-spin transitions.}

\label{tableNparams}
\end{table}

\begin{figure}
\centering{}\includegraphics[width=1\columnwidth]{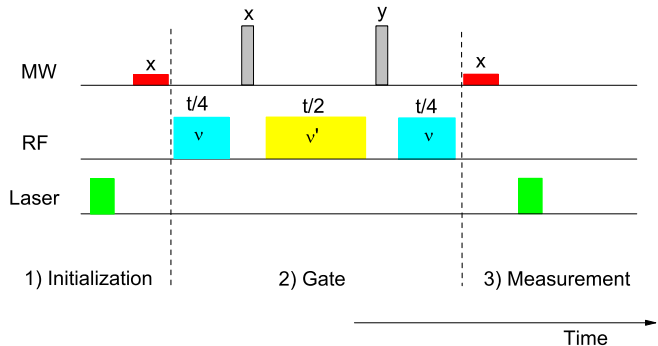} \protect\protect\protect\protect\protect\protect\protect\protect\protect\caption{(color online). Pulse sequence and experimental procedure. The microwave
(MW) pulses are resonant  with the transition $|0\rangle_{e}|0\rangle_{n}\leftrightarrow|-1\rangle_{e}|0\rangle_{n}$.
In step 1) and 3), the MW pulses have the same Rabi frequency of about
$0.28$ MHz and the same flip angle. These pulses work as transition
selective pulses. The DD pulses in step 2) are hard pulses with Rabi
frequency of about $11$ MHz. The frequencies of the radio frequency
(RF) pulses are indicated in the rectangles, where $\nu=\nu_{2}$,
$\nu'=\nu_{1}$ for a CR gate with control state $|1\rangle$ or $\nu=\nu_{1}$,
$\nu'=\nu_{2}$ for the gate with control state $|0\rangle$. \label{figEnergysub}}
\end{figure}

%%%%%%----------------------------------------------------------------------------

In the experiment, we used a $^{12}$C enriched diamond as the sample,
so that decoherence due to $^{13}$C nuclear spins is small. The experiment
was performed at room temperature. Fig. \ref{figEnergysub} shows
the experimental procedure and the pulse sequence. Since the electron
spin decoheres on a much shorter timescale than the nuclear spin,
it is sufficient to apply DD to the electron spin transitions (qubit
1). This is done in parallel to the controlled rotation of qubit 2.
The controlled rotation of qubit 2 is interrupted during the application
of the DD pulses, which are several orders of magnitude shorter than
the duration of the nuclear spin gate operations (10 ns vs. $>100$
$\mathrm{\mu s}$). The DD pulses are hard $\pi$-pulses around the
$x$- or $y$- axis, as indicated above the rectangles in Fig. \ref{figEnergysub},
with a Rabi frequency of about $11$ MHz.  For this demonstration
experiment, we used only two DD pulses. To make it symmetric with
respect to the initial condition, we alternated the rotation axes
between $x$ and $y$. To turn this into a cyclic operation (generating
a unit operation), it has to be combined with a rotation around the
$z$-axis, which we implemented as a phase shift of the subsequent
pulses.

In step 1), we prepared the input state for the CR gate. The laser
pulse initializes the electronic spin in its ground state $m_{s}=0$,
while the nuclear spin is unpolarized. The first microwave (MW) pulse
drives selectively the $|00\rangle\leftrightarrow|10\rangle$ transition,
with a $\pi/2$ flip angle around the $x$-axis. The resulting input
state is
\begin{equation}
\rho_{in}=\frac{1}{4}(|00\rangle-i|10\rangle)(\langle00|+i\langle10|)+\frac{1}{2}|01\rangle\langle01|,\label{inputstate1}
\end{equation}
written in the computation basis (\ref{eq:CompBase}) but omitting
the index. In this state, the control qubit is in a superposition
of $|0\rangle$ and $|1\rangle$. The CR gate drives the transition
labelled $\nu_{1}$ in Fig. \ref{figlevels} if the control condition
is 0 and the transition labelled $\nu_{2}$ if the control condition
is 1.

For the readout, we measure the population of the $|0\rangle$
state of the electron spin. To measure the effect of the RF pulse,
which affects only the nuclear spin, we therefore need to transfer
the information from the nuclear to the electron spin system. For
this purpose, we use again a selective MW pulse that
 rotates the $|00\rangle\leftrightarrow|10\rangle$ transition  around
the $x$-axis, by a flip angle $-\alpha$. As a function of the rotation
angle $\theta$ of the CR gate, the expected signal (i.e. the occupation
probability for the $|0\rangle$-state in the computational space)
is
\begin{equation}
s=\frac{1}{8}[1+\cos\frac{\theta}{2}]^{2}+\frac{1}{2}=\frac{1}{8}[1+2\cos\frac{\theta}{2}+\cos^{2}\frac{\theta}{2}]+\frac{1}{2}\label{pout2}
\end{equation}
for an ideal gate. In the case of an unprotected gate, the environmental
perturbation causes dephasing of the electronic spin state. For a
Markovian environment, the off-diagonal elements of the electron spin
qubit decay as $e^{-t/T_{2}}$. This changes the detected signal to
\begin{equation}
s_{T_{2}}=\frac{1}{8}[1+2\cos\frac{\theta}{2}e^{-t/T_{2}}+\cos^{2}\frac{\theta}{2}]+\frac{1}{2}.\label{pout2dec}
\end{equation}

%%%%-----------------------------------------------------
\begin{figure}
\centering{}\includegraphics[width=1\columnwidth]{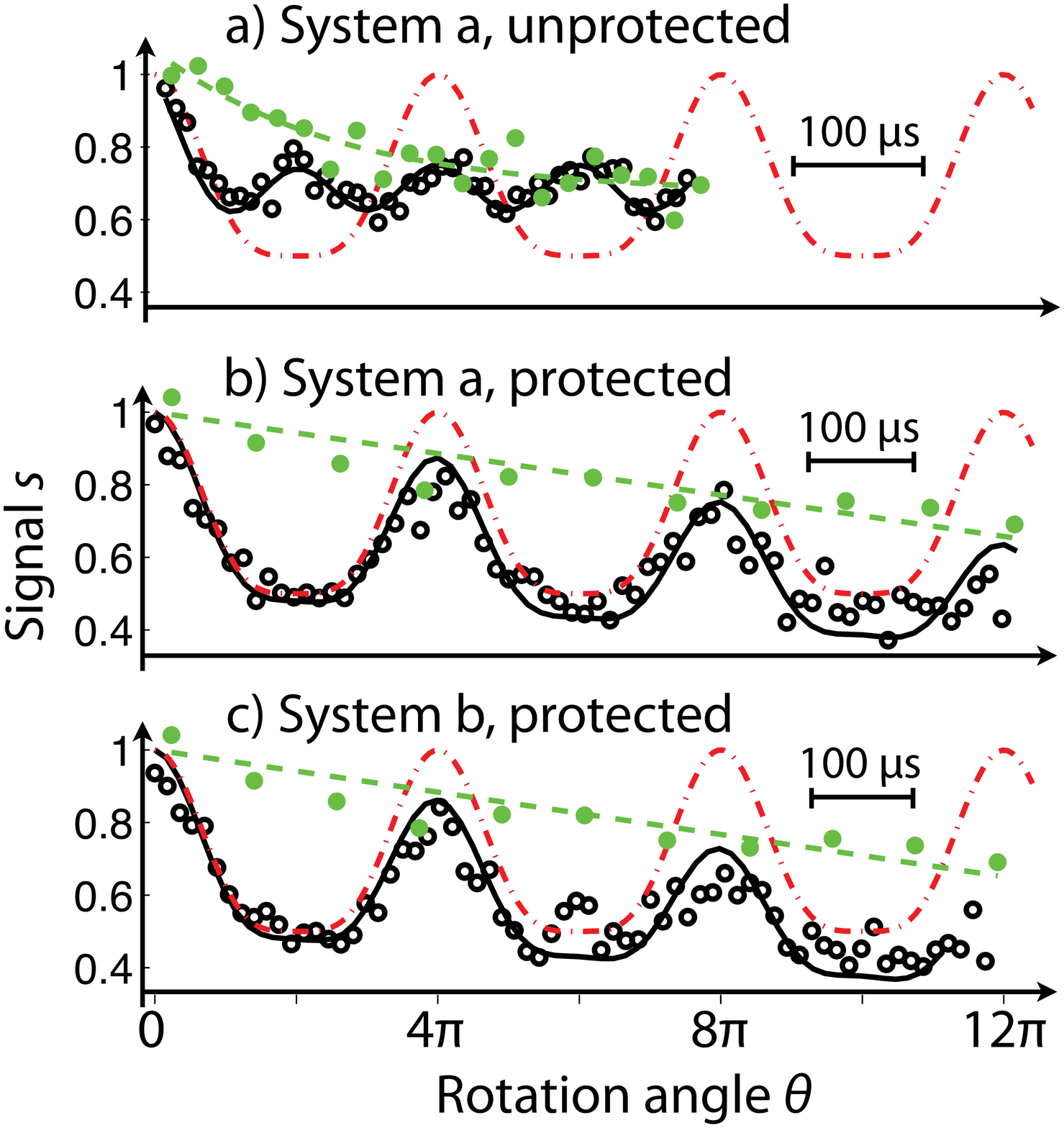}
\protect\protect\protect\protect\protect\protect\protect\protect\protect\caption{(color
online). Experiment results for implementation of the CR gates
starting with the input state $\rho_{in}$. The results for the
unprotected gate in system $a$ and the protected gate in systems
$a$ and $b$ are shown in figure (a-c), respectively. The
theoretical expectations for ideal gates are shown as red
dash-dotted curves, and the experimental data as empty circles.
The filled circles show the decay of the input state, without CR
gate and without DD pulses in figure (a), and with the DD pulses
in figures (b-c). The dashed curves show the result of a linear
fit. In figure (a), the solid curve shows the result of a
simulation with the measured dephasing time. In figure (b-c) the
solid curves show the result of a fit to the function
$s_{r}=(1-\kappa t)s_{T_{2}}$, where $\kappa$ is an empirical
decay rate. \label{figresmain4} }
\end{figure}

We implemented the unprotected gate by setting $\nu'=\nu$ and switching
off the DD pulses in the sequence in Fig. \ref{figEnergysub}. Fig.
\ref{figresmain4} (a) illustrates the resulting signal when the unprotected
CR gate was implemented with control state $|0\rangle$ in system
$a$. The horizontal axis gives the rotation angle of the CR gate
$\theta=\Omega_{1}t$, where $\Omega_{1}$ is the Rabi frequency of
the nuclear spin transition. While the ideal gate generates a signal
that is periodic in $\theta$ with a period of $4\pi$, the dephasing
of the electron spin eliminates this signal component over a time
scale of the order of the decay time $T_{2}\approx34\,\mathrm{\mu s}$
of the input state without DD pulses. The remaining signal component
is periodic with 2$\pi$, in agreement with Eq. (\ref{pout2dec}).
The measured data for the CR gate are marked as empty circles, and
the theoretical expectation $s$ for the ideal gate is shown as the
dash-dotted curve. The solid curve shows the function $s_{T_{2}}$
from Eq. (\ref{pout2dec}) and is in good agreement with the experimental
data. Clearly, the dephasing time in this system is too short to allow
the implementation of a nontrivial gate, such as CNOT, with sufficient
precision. In addition, we measured the decay of the initial state
$\rho_{in}$ due to relaxation, without applying a gate operation.
The resulting data points are shown as filled green circles.

Figs. \ref{figresmain4} (b-c), show the corresponding data obtained
by the protected CR gate in systems $a$ and $b$. The total rotation
angle is $\theta=(\Omega_{1}+\Omega_{2})t/2$. To evaluate the performance
of the gate, we first measured the decay of the input state $\rho_{in}$
protected by the DD pulses. The results are shown in Fig. \ref{figresmain4}
(b-c) as filled green circles. Over the measured time scale, the experimental
data points decay with a time constant of $1/\kappa\approx2.4$ ms,
mainly from longitudinal relaxation ($T_{1}$) and experimental errors.
For the protected gate, we used the same decay function and superimposed
it over the experimental data points in figures (b-c). The experimental
data points can then be fitted with $s_{r}=(1-\kappa t)s_{T_{2}}$,
even for times much longer than the dephasing time scale in the unprotected
gate shown in figure Figs. \ref{figresmain4} (a), using pure depahsing
times $T_{2}$ of 4 ms.

\begin{figure}[t]
\centering{}\includegraphics[width=1\columnwidth]{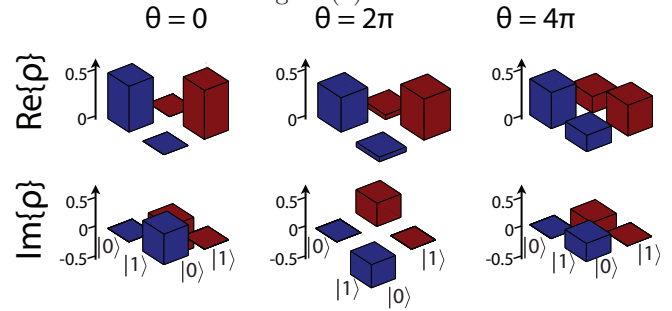} \protect\protect\protect\protect\caption{(color online). Results for state tomography for $\theta=0$, $2\pi$
and $4\pi$ shown as columns 1 - 3, respectively, where the first
and second rows show the real and imaginary parts. In theory, density
matrices are $(E+\sigma_{y})/2$, $(E-\sigma_{y})/2$ and $(E+\sigma_{y})/2$
for $\theta=0$, $2\pi$ and $4\pi$, respectively, where $E$ denotes
the identity operator. \label{figrestomo}}
\end{figure}

A precise check of the performance of the gate operation could be
performed by quantum process tomo\-graphy. However, this requires
the application of SWAP operations consisting of the CR gates that
we are evaluating. We therefore only checked the state of the control
qubit after the gate operation by quantum state tomography. The CR
gate is an entangling gate. However, when the rotation angle $\theta$
is an even multiple of  $2\pi$, the CR gate (\ref{UCRmat}) becomes
the identity or phase gate $e^{i(\pi/2)\sigma_{z}}$ for qubit 1,
the electronic qubit can be separated from the nuclear spin. At these
points, we can reconstruct the state of the electronic qubit by treating
it as a single qubit. We  implemented quantum state tomography for
$\theta=0$, $2\pi$ and $4\pi$. To reconstruct each state, we used
five measurements with different MW pulses in step 3): no pulse, and
pulse along $x$, $-x$, $y$ and $-y$, respectively. Fig. \ref{figrestomo}
illustrates the results. One can clearly observe the effect of the
$z$-rotation i.e., phase flip, in the results for $\theta=2\pi$.

We compared the experimentally measured states with the ideal
states by calculating the fidelity
$F=|Tr\{\rho_{exp}\rho_{ideal}\}|$ and obtained the values
$F=0.90$, $0.66$ and $0.57$ for $\theta=0$, $2\pi$ and $4\pi$,
respectively. We can attribute the difference between the
experimental and ideal results mostly to longitudinal relaxation,
which was measured to be  $\approx3.5$ ms by the standard MW pulse
sequences  and cannot be reversed by dynamical decoupling, and to
experimental imperfections.

In addition to the experiments discussed here, we applied the protected
 CR gate also to other input states and found good agreement with
the theoretical predictions. In those cases, where the initial state
does not contain coherence of the electron spin, the unprotected gate
also works relatively well \cite{PhysRevA.87.012301,dobrovitski2012}.
The resulting evolution is described by Eq. (\ref{pout2dec}) with
$T_{2}\rightarrow0$, and it corresponds to the results shown in Fig.
\ref{figresmain4} (a) for $\theta>2\pi$.

\textit{--Conclusion.} We have demonstrated the protection of 2-qubit
gate operations by dynamical decoupling by combining the control field
with dynamical decoupling of the control qubit, which otherwise undergoes
rapid dephasing. The protocol can be generalized for multiple-qubit
gates, e.g., to NV centers in natural abundance diamond samples, where
$^{13}$C nuclear spins can act as qubits. The same strategy and techniques
should be helpful for implementing quantum control in other quantum
systems, in particular for hybrid systems. In future work, we plan
to generalize this scheme  and use more advanced DD sequences, such
as the KDD sequence \cite{suter2011}.

{\it Acknowledgments.--} JZ acknowledges F. D. Brandao for helpful
discussions. This work is supported the DFG through Su 192/19-2.

%---------------------------------------------

% \bibliographystyle{apsrev}
%\bibliography{NVC_CNOT}

\end{document}